# Multiscale Modeling of Semimetal Contact to Two-Dimensional Transition Metal Dichalcogenide Semiconductor

Tong Wu and Jing Guo[a)]

*Department of Electrical & Computer Engineering, University of Florida, Gainesville, FL, 32611, USA*
*[a)]Author to whom correspondence shall be addressed: guoj@ufl.edu*

*Abstract*— A multiscale simulation approach is developed to simulate the contact transport properties between semimetal to a monolayer two-dimensional (2D) transition metal dichalcogenide (TMDC) semiconductor. The results elucidate the mechanisms for low contact resistance between semimetal and TMDC semiconductor contacts from a quantum transport perspective. The simulation results compare favorably with recent experiments. Furthermore, the results show that the contact resistance of a Bismuth-MoS$_2$ contact can be further reduced by engineering the dielectric environment and doping the TMDC material to $< 100 \, \Omega \cdot \mu m$. The quantum transport simulation indicates the possibility to achieve an ultrashort contact transfer length of ~1 nm, which can allow aggressive scaling of the contact size.

Reducing contact resistance to two-dimensional (2D) transition metal dichalcogenide (TMDC) semiconductors is of crucial importance for device technologies based on 2D materials.[1] Recent experiments have demonstrated low contact resistance values to MoS$_2$. A contact resistance value of ~123 $\Omega \cdot \mu m$ was reported for a contact between semimetal bismuth (Bi) and MoS$_2$,[2] and a value of ~660 $\Omega \cdot \mu m$ was reported for a contact between antimony semimetal and MoS$_2$[3]. The contact stacks semimetal on top of the 2D TMDC material, which has a schematic structure as shown in Fig. 1(a) *Ab initio* density-functional theory (DFT) simulations shows that the semimetal Fermi level is aligned near the conduction band edge of MoS$_2$, which facilitates low contact resistance.[2,3] Direct evaluation of contact resistance values, however, requires transport and device studies. For contacts to 2D semiconductors, interface states and atomistic scale features, self-consistent electrostatics, and quantum transport effects can all affect contact resistance.[4] Compared to metal contacts to bulk semiconductors, the charge transport physics in contacts to 2D semiconductors can differ considerably, due to reduced semiconductor dimensionality.[5, 6]

To bridge the importance of atomistic scale features at the contact interface and the need to model an extended length scale in the horizontal transport direction, we developed a multiscale simulation approach to model quantum transport properties of semimetal-TMDC contacts. We investigated the role of metal-induced gap states (MIGS), compared the modeled contact resistance to the experimental value, and assessed the potential to further reduce the contact resistance. The modeled contact resistance is in good agreement with experiments. The contact transfer length is also investigated by quantum transport simulations, which illustrates the potential to achieve aggressive scaling of the contact size. We focus on semi-metal contacts in this study due to its potential to achieve low contact resistance.

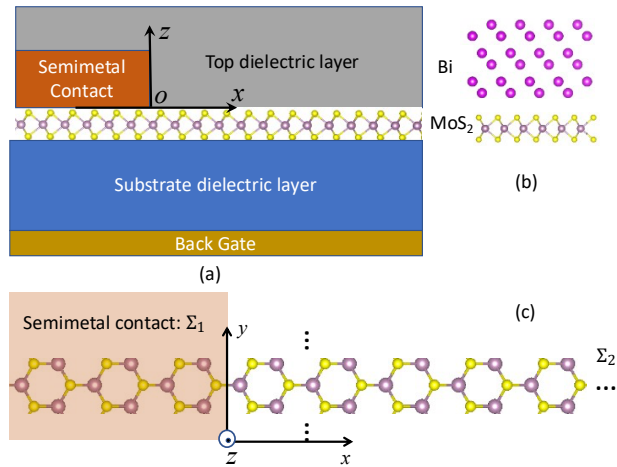

FIG. 1. (a) Schematic side view of the modeled contact structure. (b) Side view of the atomistic structure for *ab initio* DFT simulations of the Bi-monolayer MoS$_2$ contact. (c) Top view of the contact modeled by atomistic tight binding NEGF simulations. The coordinate system is shown, where $x$ is the transport direction, $y$ is the transverse direction, and $z$ is the vertical contact stacking direction.

To capture interface atomistic scale features, self-consistent electrostatics, and quantum transport properties at a larger scale for the contact structure as shown in Fig. 1, a multiscale simulation approach, which includes atomistic tight-binding (TB) quantum transport simulations with contact self-energy determined from *ab initio* DFT simulations, and self-consistent electrostatics determined by Poisson equation, is developed as described in detail below. Compared to the self-consistent transmission line model approach[7], the multiscale modeling approach described here also treats self-consistent electrostatics. The multiscale modeling approach performs quantum transport simulation with an atomistic description of the semiconductor. It captures non-parabolic band structure, quantum effects, and can be extended to model defects and roughness by including a defect potential term in the Hamiltonian (which is not included in this study).

A TB atomistic Hamiltonian is used for transport in monolayer MoS$_2$, which results in a low-energy $E$-$k$ relation that captures coupled spin and valley physics in monolayer MoS$_2$.[8] For the bipartite lattice as shown in Fig. 1(c), an atomistic Hamiltonian with one basis orbital per site is used. The transport is along the *x*-direction, in which a real space treatment is used. In the transverse y-direction, a *k*-space treatment is used, and the physical quantities of interest are obtained by summing over the transverse *k* modes. The onsite energies of the bipartite lattice site at $x$ are $h_{A0} = E_m(x) + \Delta/2$ and $h_{B0} = E_m(x) - \Delta/2$, and the first nearest neighbor (n.n.) TB parameter is $t$, where for $MoS_2$, $\Delta = 1.66$ eV and $t = -1.10$ eV. [8] To treat spin splitting in the valence band, a $2^{nd}$ n.n. interaction in the B sublattice is introduced, with an imaginary TB parameter, $h_{BB2} = icS_z t_2$, where $i$ is the imaginary unit, $c = \pm 1$ is the sign parameter depends on whether the bond makes a left or right turn to the $2^{nd}$ n.n., $S_z = \pm 1$ is the *z* spin, $t_2 = \lambda/(3\sqrt{3})$ and $2\lambda = 0.15$ eV is the valence band spin splitting of monolayer $MoS_2$.[8]

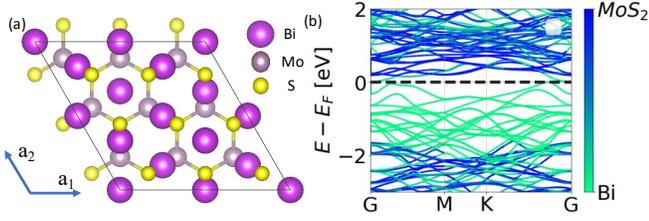

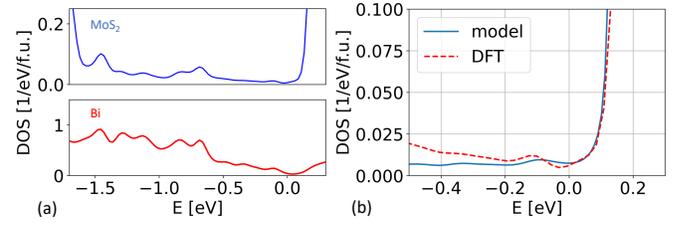

FIG. 2. *Ab initio* supercell simulation of the interface: (a) top view of the simulated supercell structure. (b) simulated supercell band structure. The $MoS_2$ and Bi contributions to the bands are color coded as shown in the color bar. The Fermi level is denoted as the dashed line at $E$=0 eV.

FIG. 3. (a) Density of states of the $MoS_2$ layer (top) and Bi interface atom (bottom) from DFT simulations. (b) Comparison of the density of states of the $MoS_2$ layer from the DFT simulations (dashed red line) to that of the TB model with broadening due to coupling to the Bi contact (solid blue line).

Based on the TB Hamiltonian, quantum transport simulations were performed by using the non-equilibrium Green's function (NEGF) method. The right-side contact is a semi-infinite contact described by a contact self-energy of $\Sigma_2$, as shown in Fig. 1(c). The atoms in the metal-contact-covered region are coupled to the metal contact, which is described by a phenomenological self-energy $\Sigma_1$. A multiscale method is developed to extract the self energy $\Sigma_1$ for the TB model from the *ab initio* DFT simulations as described below.

To capture the contact interface atomistic features from first principles, *ab initio* DFT simulations were performed by using the Vienna *ab initio* simulation package (VASP)[9] for a Bi-monolayer $MoS_2$ interface as shown in Fig. 1(b). A supercell structure is constructed for interface simulations, as shown in Fig. 2(a). The calculation was performed by using the projector augmented-wave (PAW) pseudopotential, and the generalized gradient approximation (GGA) method was used with a Perdew–Burke–Ernzerhof (PBE) exchange-correlation functional. The Brillouin-zone was sampled by a 25×25×1 Monkhorst-Pack scheme for calculating the atom-resolved density of states (DOS)[10]. The supercell structure of Bi-$MoS_2$ contact is formed by stacking the 2×2 Bi lattice on the 3×3 monolayer 2H-$MoS_2$ lattice in the same direction and with a separation of 3.5 Å, as shown in Fig. 2(a). The simulated band structure of the supercell Bi-$MoS_2$ stack, as shown in Fig. 2(b), indicates that the Fermi level is placed near the conduction band edge.

The multiscale simulation extracts the self-energy parameters of the semimetal contact from *ab initio* simulations. The contact self-energy can be expressed as,

$$\Sigma = \tau g_s \tau^+, \tag{1}$$

where $g_s$ is the surface Green's function of the contact, and $\tau$ is the coupling between the channel and contact. The contact broadening is

$$\Gamma = i(\Sigma - \Sigma^+) = \tau A_s \tau^+ = 2\pi\tau D_s \tau^+, \tag{2}$$

where $A_s = i(g_s - g_s^+) = 2\pi D_s$ is the contact surface spectral density and $D_s$ is the surface DOS of the Bi contact. The contact resistance is determined by transport in a small energy range near the Fermi level and conduction band edge, and we assume $\tau$ is constant in this energy range. We consider on-site broadening from the metal contact only, and assume $\Gamma$ is diagonal, which the diagonal entry being

$$\gamma(E) = 2\pi\tau_0^2 D_{Bi}(E). \tag{3}$$

The energy dependence of the contact broadening is determined by that of the DOS of the Bi contact atom $D_{Bi}(E)$, and $\tau_0$ is extracted by fitting the TB results to the *ab initio* DFT simulation results of the interface as described below.

The band profile of the contact structure in Fig. 1(a) is influenced by self-consistent electrostatics and MIGS. At the metal-semiconductor interface, the metal states penetrate into the semiconductor, which results in an exponentially decaying density of states in the band gap in the semiconductor called MIGS[11,12]. We treat the MIGS following Ref. [13], which models an interface dipole induced by MIGS by a charge,

$$\sigma(x) = \begin{cases} D_0(E_N - E_F), & if \ x < 0 \\ D_0(E_N - E_F)e^{-qx}, & if \ x \geq 0 \end{cases}, \tag{4}$$

where the horizontal position $x = 0$ is defined at the boundary between the metal covered and the extension regions as shown in Fig. 1. Eq (4) provides a concise phenomenological description of the MIGS [11,12,13]. The value of $1/q$ characterizes the average characteristic decay distance of the MIGS into semiconductor. A typical value of $q \sim 1 \ nm^{-1}$ to $2 \ nm^{-1}$ has been previously used. Here we use a value of $q \approx 2 \ nm^{-1}$, which is further cross-validated by NEGF simulations later. The values of $D_0$ and $E_N$ depend on atomistic details at the interface and have uncertainties. We test the charge neutrality level in the range of $E_N = E_c - 0.15$ eV to $E_N = E_c$, as *ab initio* simulations indicate that the Fermi level is near the conduction band edge for the Bi-$MoS_2$ top contacts, which is shown in Fig. 2 and Fig. 3(a). An average value of $D_0 \approx 0.01 \ /eV/f.u.$ is extracted from *ab initio* interface simulations in the energy range near $E_N$, where a formula unit ( $f.u.$ ) of the semiconductor is $MoS_2$, as shown in Fig. 3(b). Fig. 3(b) also shows the comparison of the density of states from the *ab initio* simulation to the TB contact self-energy mode in Eq. (3) where the extracted value $\tau_0 \approx 0.18$ eV.

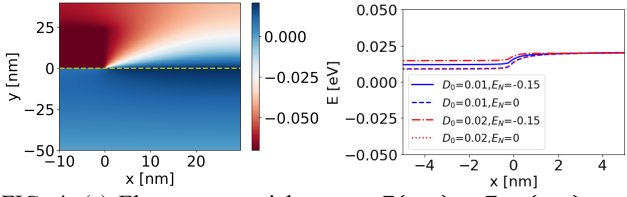

FIG. 4. (a) Electron potential energy, $E(x, y) = E_{vac}(x, y) - \chi_s$, computed as the vacuum level minus the semiconductor affinity. The bottom oxide thickness is 50 nm with a dielectric constant of 4, the Bi contact length is 10 nm and the height is 25 nm (in $-10\ nm < x < 0$ and $0 < y < 25\ nm$). The dashed line shows the position of $MoS_2$ layer at $y = 0$. (b) Zoomed-in conduction band edge vs. lateral position $x$ with different charge neutrality levels of $E_N = 0$ and $E_N = -0.15$ eV and MIGS values of $D_0 = 0.01/eV/f.u.$, and $D_0 = 0.02/eV/f.u..$ The modeled device structure is shown in Fig. 1(a).

To obtain the electrostatic potential and band profile, a numerical two-dimensional non-linear Poisson solver[14] in the contact cross section as shown in Fig. 1(a) is solved with the equilibrium carrier statistics in the presence of MIGS states. The total charge density is determined by the sum of the charge due to the MIGS states and band electrons, whose density is

$$n_{2D} = N_{C2D} \mathcal{F}_0((E_F - E_C(x))/k_B T), \quad (5)$$

where $N_{C2D} = g_s g_v (\frac{m_{eff} k_B T}{2\pi \hbar^2})$ , $\mathcal{F}_0(x) = \ln(1 + \exp(x))$ is Fermi integral to the zeroth order, $E_F = 0$ is the equilibrium Fermi level, $E_C(x)$ is the conduction band edge, and $k_B$ is the Boltzmann constant. For $MoS_2$, we treat the conduction band with the spin degeneracy factor of $g_s = 2$, valley degeneracy factor of $g_v = 2$, and an electron effective mass of $m_{eff} = 0.57 m_0$[15], where $m_0$ is the free electron mass.

The vacuum energy level, which is continuous in space, is solved based on the non-linear Poisson equation. In the Bi contact region, the Dirichlet boundary condition is used for the vacuum level, $E_{vac,Bi} = E_F + \phi_{Bi}$, where the Bi work function $\phi_{Bi}$ was experimentally characterized to be between 4.22 eV and 4.25 eV[16], and a value of $\phi_{Bi} = 4.23$ is used here. A Neumann boundary condition is used for the top, left, and right boundary. The electron affinity of the semiconductor monolayer 2H-$MoS_2$ is $\chi_S \approx 4.3$ eV [17]. A bottom oxide thickness of 50 nm and dielectric constant of 4 are assumed, and a Dirichlet boundary condition is used for the bottom contact with an applied voltage of $V_{GB} = 0$ and $\phi_{BG} = 4.3$ eV. The results are insensitive to further increase of the bottom oxide thickness. After the vacuum energy is solved, the conduction band edge in the semiconductor channel is obtained as $E_C(x) = E_{vac}(x, y = y_0) - \chi_s$, where $y_0$ is the vertical position of the $MoS_2$ layer.

The simulated electron potential is shown in Fig. 4(a), where the $MoS_2$ layer is located at $y_0 = 0$. The Bi contact produces a fringing electric field to the $MoS_2$ layer. In order to investigate the impact of the MIGS parameters on the band profile, the conduction band edge profiles with different charge neutrality levels and $D_0$ values are tested. The results show that the uncertainty of the MIGS parameters only has a small impact on the band profile, especially in the extension region of $x > 0$, as shown in Fig. 4(b).

We next discuss the results from multiscale quantum transport simulation, which has an atomistic Hamiltonian description for the 2D semiconductor, and thereby, captures non-parabolic energy dispersion of the 2D material. The contact resistance of the monolayer 2D semiconductor material can be sensitive to its doping and electrostatic environment. The monolayer semiconductor material is sandwiched between the substrate and top dielectric layers. The monolayer semiconductor can be doped either chemically or electrostatically. A fringing electric field between the metal contact and the monolayer semiconductor can influence the band profile near the contact. Doping and electrostatic environment can impact the potential profile at the contact, and thereby, the contact resistance.

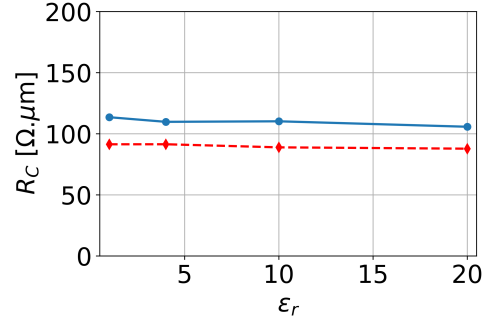

FIG. 5. Simulated contact resistance vs. the dielectric constant of top insulator for the contact structure as shown in Fig. 1. The $MoS_2$ doping density is $N_D = 4 \times 10^{12}\ cm^{-2}$ (solid line) and $5 \times 10^{12}\ cm^{-2}$ (dashed line).

We investigate the effect of the semiconductor doping density and the dielectric constant of the top dielectric layer on the contact resistance in Fig. 5. The relative dielectric constant of the substrate is fixed at $\kappa_{sub} = 4$ and that of the top dielectric layer is varied from $\kappa_{top} = 1$, which is typical for a back-gated device structure, to a value of $\kappa_{top} = 20$ for a high-$\kappa$ $HfO_2$ top gate insulator. The metal contact height is $t_{metal} = 25\ nm$, which is sufficient, i.e., the further increase does not impact the contact resistance. The results in Fig. 5 show that the contact resistance reduces slightly as the top dielectric constant increases, due to a larger fringing field and stronger electrostatic coupling between the sidewall of metal contact to the $MoS_2$ monolayer.

As the n-type doping density in the monolayer semiconductor increases from $N_D = 4 \times 10^{12}\ cm^{-2}$ to $5 \times 10^{12}\ cm^{-2}$, the conduction band in the semiconductor moves

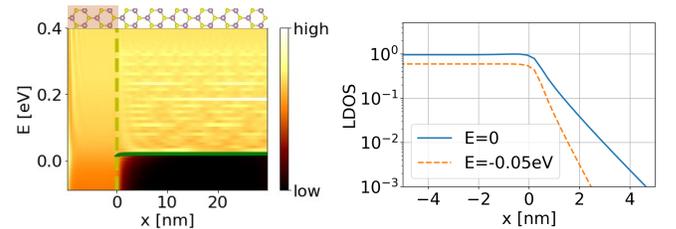

FIG. 6. (a) Pseudo color plot of LDOS of the $MoS_2$ layer. The vertical dashed line at $x = 0$ shows the boundary between the semimetal-covered ($x < 0$) and extended ($x > 0$) regions, and the solid line shows the conduction band edge. (b) LDOS vs. the horizontal position x at energy $E = 0$. (b) Normalized LDOS vs. $x$ at energy $E = 0$ (solid) and $E = -0.05$ eV (dashed) The exponential decay of the MIGS states in semiconductor has slopes of $q \approx 1.4\ nm^{-1}$ at $E = 0$ and $q \approx 2.6\ nm^{-1}$ at $E = -0.05$ eV.

closer to the Fermi level. As a result, the contact resistance reduces.

We next compare the simulated contact resistance to the experimental value. The experimental result reported in Ref. [2] has air on top of the MoS$_2$ layer which corresponds to $\kappa_{top} \approx 1$, and a SiO$_2$ dielectric substrate with $\kappa_{sub} \approx 4$. The experimentally reported contact resistance is $R_C \approx 123 \, \Omega \cdot \mu m$. For the simulated device structure of $\kappa_{top} \approx 1$ and $\kappa_{sub} \approx 4$, the modeled contact resistance value depends on the doping density. As the doping density varies between $N_D = 4 \times 10^{12} \, cm^{-2}$ to $5 \times 10^{12} \, cm^{-2}$, the simulated contact resistance varies from $115 \, \Omega \cdot \mu m$ to $91 \, \Omega \cdot \mu m$. As the exact doping density due to electrostatic gating and intentional or unintentional doping of the semiconductor is unavailable, and uncertainties exist in material and structural parameters, the comparison is semi-quantitative. The results also show that a low contact resistance of $R_C \approx 91 \, \Omega \cdot \mu m$ at a semiconductor doping density of $N_D = 5 \times 10^{12} \, cm^{-2}$ and top layer dielectric constant of $\kappa_{top} = 1$. The value reduces to $R_C < 90 \, \Omega \cdot \mu m$ if the doping density or $\kappa_{top}$ increases.

To investigate the characteristics of MIGS, we plot the simulated local density of states (LDOS), $LDOS(x, E)$, in Fig. 6 for the modeled contact as shown in Fig. 1. The contact edge horizontal position $x = 0$ is shown by a vertical dashed line in Fig. 6(a), which is the boundary between the contact-covered ($x < 0$), and the extension ($x > 0$) regions. The conduction band edge $E_C(x)$ in the extension region is also shown. The results indicate a finite MIGS in the entire bandgap of the contact-covered region, as well as decaying MIGS stages beyond $x > 0$. Fig. 6(b) plots LDOS vs. position $x$ at two different energies, which confirms the qualitative feature of MIGS as described by Eq. (1), with an extracted value of the exponentially decaying rate of $q \approx 1.4 \, nm^{-1}$ at $E = 0$ and $q \approx 2.6 \, nm^{-1}$ at $E = -0.05 \, eV$. The exponential decay reflects the quantum tunneling nature of the MIGS states.

The contact transfer length is investigated next. The left axis of Fig. 7 plots the position-resolved current density $J(x)$ from the Bi contact to the MoS$_2$ layer in the contact-covered region ($x < 0$) for the contact structure as shown in Fig. 1, fitted by an exponential relation as shown by the dashed line, $J(x)/J_0 = \exp(-|x|/L_T)$, where the normalization factor $J_0$ is the value of $J(x)$ at $x = 0$, and $L_T$ is the contact transfer length. A best fitting value of $L_T \approx 0.8 \, nm$ is extracted to ballistic NEGF simulation results as shown by symbols, which indicates the possibility of achieving a very short contact transfer length at the ballistic transport limit in MoS$_2$. In the diffusive transport limit, the contact transfer length can be computed as $L_{T,diff} = \sqrt{\rho_C/\rho_s}$, where $\rho_c$ is the contact specific resistivity, and $\rho_s$ is the sheet resistivity of the semiconductor.

By applying this relation to experimental data, a value of $L_{T,diff} \approx 3 \, nm$ is extracted. At the ballistic transport limit, there is no sheet resistivity. The contact transfer length is limited by the barrier and coupling between metal and semiconductor cross the interface. The right axis shows the percentage of the cumulative current $\gamma = \int_x^0 J(x')dx' / \int_{-\infty}^0 J(x')dx'$ in the beginning distance of $|x|$ from the contact edge. The results indicate that $\gamma \approx 98\%$ at $x = -3 \, nm$, which is the first 3 nm of the contact distance from its edge. The results indicate the potential to achieve an ultrashort contact and aggressively scale down the contact size.

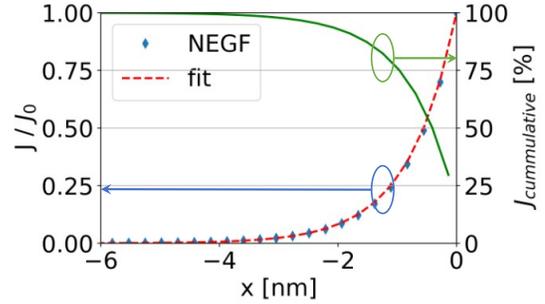

FIG. 7. Simulated normalized current density vs. position in the semimetal-covered region (left) and the percentage of the cumulative total current density from the contact boundary (right) for the modeled structure as shown in Fig. 1. The red dashed line shows exponential fitting to the current density (left). The Bi contact has a boundary at $x = 0$, and the semimetal-covered region is $x < 0$.

In summary, by developing a multiscale method that integrates *ab initio* DFT simulations with tight-binding quantum transport simulations, the contact properties of Bi-MoS$_2$ are simulated. The results illustrate the low and rapidly decaying properties of the MIGS states, and indicate the possibility to achieve a low contact resistance < 100 $\Omega \cdot \mu m$. An ultrashort contact transfer length ~1 nm is extracted for ballistic transport in MoS$_2$, which indicates the potential to achieve aggressively scaled contact size without sacrificing contact resistance. The multiscale approach developed in this work can also be applied to contacts to other 2D TMDC semiconductors.

## Acknowledgment

This work was supported by the National Science Foundation, Awards #1809770, and #1904580.

## Data availability

The data that support the findings of this study are available from the corresponding authors upon reasonable request.